\documentclass{PoS}
\usepackage{macros}
\usepackage{amsmath}
\title{On O($a^2$) effects in gradient flow observables}

\ShortTitle{On O($a^2$) effects in gradient flow observables}

\author{Alberto Ramos \\\\ PH-TH CERN, CH-1211 Geneva 23, Switzerland\\
        E-mail: \email{alberto.ramos@cern.ch}}

\author{\speaker{Stefan Sint}  \\\\
        NIC, DESY-Zeuthen, Platanenallee 6 D-15738 Zeuthen, Germany\\
        \& School of Mathematics, Trinity College, Dublin 2, Ireland\\
        E-mail: \email{sint@maths.tcd.ie}}

\abstract{In lattice gauge theories, the gradient flow has been used extensively both,
for scale setting and for defining finite volume renormalization schemes for the gauge coupling.
Unfortunately, rather large cutoff effects have been observed in some cases.
We here investigate these effects to leading order in perturbation theory,
considering various definitions of the lattice observable,
the lattice flow equation and the Yang Mills lattice action.
These considerations suggest an improved set-up for which we perform
a scaling test in the pure SU(3) gauge theory, demonstrating strongly reduced cutoff effects.
We then attempt to obtain a more complete understanding of the structure
of O($a^2$) effects by applying Symanzik's effective theory approach to the 4+1 dimensional
local field theory with flow time as the fifth dimension.
From these considerations we are led to a fully O($a^2$) improved set-up the study of
which is left to future work.
\vskip 15ex
\begin{flushleft}
DESY 14-202\\
TCDMATH 14-09\\
CERN-PH-TH-2014-214
\end{flushleft}

}

\FullConference{The 32nd International Symposium on Lattice Field Theory,\\
		23-28 June, 2014\\
		Columbia University New York, NY}

\begin{document}

\section{Yang-Mills gradient flow and running couplings}

The gradient flow~\cite{Luscher:2010iy} allows for the definition of renormalized observables
which are measurable with high statistical precision at relatively low computational
costs. One may thus hope that this will allow  for high precision in
the calibration of the lattice scale and precision studies of
the running coupling~\cite{Sommer:2014mea}.
In both cases one starts from the observable $\langle E \rangle$,
\begin{equation}
 \langle E(t,x)\rangle =-\frac12 \left\langle {\rm tr}\left(G_{\mu\nu}(t,x)G_{\mu\nu}(t,x)\right)\right\rangle,
  \qquad G_{\mu\nu} = \partial_\mu B_\nu - \partial_\nu B_\mu + [ B_\mu, B_\nu ],
  \label{eq:obs}
\end{equation}
and the gauge field $B_\mu$ at flow time $t$ is obtained as the solution
of the Yang-Mills gradient flow equation,
\begin{equation}
 \partial_t B_\mu(t,x) = - \frac{\delta S[B]}{\delta B_\mu(t,x)} = D_\nu G_{\nu\mu}(t,x) , \qquad B_\mu(0,x) = A_\mu(x),
\end{equation}
with the covariant derivative $D_\mu = \partial_\mu + [B_\mu, \cdot]$
and $A_\mu(x)$ the fundamental 4-dimensional gauge field.
In \cite{Luscher:2011bx} a perturbative all-order proof was given
that gauge invariant observables like (\ref{eq:obs}) are renormalized 
if expressed in terms of a renormalized gauge coupling, $g^2$. Given the
perturbative expansion,
$   \langle E \rangle = {\cal E}_0 g^2 + {\rm O}(g^4),$
one may thus introduce the renormalized gradient flow coupling,
\begin{equation}
   {\bar g}^2_{\rm GF}(q=1/\sqrt{8t}) =  {\cal N}^{-1} t^2 \langle E(t,x) \rangle,
   \label{eq:GFcoupling}
\end{equation}
where the constant ${\cal N}=t^2 {\cal E}_0$ ensures the standard
normalization of a coupling constant.
The requirement that ${\bar g}^2_{\rm GF}(q)$ takes a particular value can
also be seen as the implicit definition of the corresponding scale. In particular,
the scale, $t_0$, defined by~\cite{Luscher:2010iy},
\begin{equation}
   \left\{ t^2 \langle E(t)\rangle \right\}_{t=t_0} = 0.3,
\end{equation}
corresponds to a value $\bar{g}^2_{\rm GF}(q=1/\sqrt{8t_0}) = 8\pi^2/5 \approx 15.8$ (assuming $N=3$ colours).
Analogous considerations can be made in a finite space-time volume, leading
to interesting finite volume schemes for renormalized couplings.
In a symmetric (hyper-) box of physical size $L^4$
one then relates $L$ to $t$ by setting the ratio $c=\sqrt{8t}/L$
to a particular value, e.g.~$c=0.3$. Different schemes are obtained for different
boundary conditions~\cite{Fodor:2012td,Fritzsch:2013je,Ramos:2014kla,Luscher:2014kea} and
values of $c$. Here we will focus on the pure gauge theory with either
twisted periodic~\cite{Ramos:2014kla} or SF boundary conditions~\cite{Fritzsch:2013je}.

Relatively large cutoff effects have been found in $t_0$~\cite{Luscher:2010iy}
and step scaling functions of finite volume flow couplings related
to $E(t,x)$~(e.g.~\cite{Fodor:2012td}). We here present a systematic study
of O($a^2$) effects in ${\cal E}_0$,
for various lattice gauge actions, lattice definitions of the observable $E(t,x)$
and of the lattice gradient flow equation. We start with the infinite volume case,
then move to finite volume, which leads to further improvement conditions.
We then sketch how full O($a^2$) improvement \`a la Symanzik \cite{Symanzik:1983dc,Symanzik:1983gh,Luscher:1984xn}
can be achieved, based on the local $4+1$ dimensional field theory for gradient flow observables~\cite{Luscher:2011bx}.

\section{Infinite lattice}

To study the expectation value $\langle E(t,x)\rangle$ in perturbation theory on the infinite lattice
we must choose a discretization of the observable $E(t,x)$, a lattice action
and a lattice version of the flow equation. At leading perturbative order
one needs to expand $E(t,x)$  to second order in the $B_\mu$ fields, relate
these to the fundamental gauge field $A_\mu$ via the linearized lattice flow equation, and calculate the
expectation value of the two $A_\mu$ fields, i.e.~the gauge field propagator, for the chosen lattice action.
The propagator is the inverse kernel for the action when expanded to second order in
the gauge fields\footnote{We decompose e.g.~$A_\mu(x) = A_\mu^b(x) T^b$ with summation
over $b=1,\ldots,N^2-1$ understood and anti-hermitian generators $T^b$, normalized by $\tr(T^aT^b)=-\delta^{ab}/2$.},
\begin{equation}
  S = \frac12 \int_{-\pi/a}^{\pi/a} \rmd^4 p\, \tilde{A}^b_\mu(-p) K_{\mu\nu}(p,\lambda) \tilde{A}^b_\nu(p) + \rmO(A^3),
\end{equation}
where $\lambda$ is the gauge fixing parameter required for $K$ to be invertible. 
Similarly, the gradient flow equation on the lattice,
\begin{equation}
  \partial_t V_\mu(t,x) = - \partial_{x,\mu}\left(g_0^2 S_{\rm lat}[V]\right) V_\mu(t,x),\qquad V_\mu(0,x)=U_\mu(x),
\end{equation}
with links parameterized by $V_\mu(t,x) = \exp\{a B_\mu(t,x)\}$,
and expanded to first order in $B_\mu$ can be parameterized by the quadratic part of the
corresponding lattice action, and thus by another kernel $K_{\mu\nu}(p,\alpha)$,
with gauge parameter $\alpha$.
Finally, since the observable (\ref{eq:obs}) has the form of an action density, 
another free action kernel can be used, however, this time
without a  gauge fixing term. Hence, we have 3 kernels
$K_{\mu\nu}$ for action (${\rm a}$), observable (${\rm o}$) and flow (${\rm f}$).
We have considered generic lattice actions with all 4- and 6-link Wilson loops, parameterized by
coefficients $c_i$, $i=0..3$ satisfying the normalization condition\footnote{In
the literature $c_2$ and $c_3$ are sometimes interchanged.
We here follow the convention where $c_2$ multiplies the ``bent rectangles" or "chairs", and $c_3$
the ``parallelograms".}, $c_0+8c_1 +16 c_2 + 8 c_3 =1$~\cite{Luscher:1984xn}.
This includes the Wilson plaquette action ($c_0=1$, $c_1=c_2=c_3=0$)  and the tree-level O($a^2$)
improved L\"uscher-Weisz action~\cite{Luscher:1984xn} ($c_0=5/3$, $c_1=-1/12$, $c_2=c_3=0$).
For the observable we use either the corresponding lattice action densities
or the action density obtained with the clover definition of $G_{\mu\nu}(t,x)$.
For instance, the kernel corresponding to the Wilson plaquette action is given by,
\begin{equation}
  K_{\mu\nu}(p,\lambda) = \hat{p}^2\delta_{\mu\nu} + (\lambda-1) \hat{p}_\mu \hat{p}_\nu,
  \qquad \hat{p}_\mu = \frac{2}{a}\sin\left(\frac12 ap_\mu\right),
\end{equation}
with continuum limit
\begin{equation}
  K^{\rm cont}_{\mu\nu}(p,\lambda) = p^2\delta_{\mu\nu} + (\lambda-1) p_\mu p_\nu,
\end{equation}
and a small $a$-expansion of the form
\begin{equation}
  K_{\mu\nu}(p,\lambda) = K^{\rm cont}_{\mu\nu}(p,\lambda) + a^2 R_{\mu\nu}(p,\lambda) + \rmO(a^4),
\end{equation}
where
\begin{equation}
   R_{\mu\nu}(p,\lambda) = -\frac{1}{12} p^4 \delta_{\mu\nu} 
   + \frac{1}{24} (\lambda-1) 	p_\mu p_\nu (p_\mu^2+p_\nu^2).
\end{equation}
Proceeding analogously for all kernels, we obtain the master equation 
\begin{equation}
 t^2{\cal E}_0 =\frac12(N^2-1)\int_{-\pi/a}^{\pi/a} \dfrac{\rmd^4 p}{(2\pi)^{4}}
 \tr \left[K^{(\rm o)}(p,0)\rme^{-t K^{(\rm f)}(p,\alpha)} K^{(\rm a)}(p,\lambda)^{-1}\rme^{-tK^{(\rm f)}(p,\alpha)}\right].
\end{equation}
Evaluation of the trace over Lorentz indices and extension of the  momentum integrals to infinity leads to
\begin{equation}
 t^2{\cal E}_0 = \dfrac{3(N^2-1)}{128\pi^2}\biggl\{ 1 +   \dfrac{a^2}{t}
\biggl[\left(d^{(\rm o)}_1 - d^{(\rm a)}_1\right) J_{4,-2} + \left(d^{(\rm o)}_2 - d^{(\rm a)}_2\right)J_{2,0}
-2d^{(\rm f)}_1 J_{4,0} -2 d^{(\rm f)}_2 J_{2,2}\biggr]\biggr\}
\label{eq:oaeffects}
\end{equation}
where corrections are O($a^4$) and
\begin{equation}
 J_{n,m} = \dfrac{t^{(m+n)/2}\int_{-\infty}^\infty \rmd^4p\, \rme^{-2tp^2} p^n p^m}
                 {\int_{-\infty}^\infty \rmd^4p\, \rme^{-2tp^2}},
 \qquad p^n\Bigl\vert_{n=2,4,...} = \sum_\mu p_\mu^n, \qquad p^{-n} = 1/p^n.
\end{equation}
All momentum integrals can be evaluated analytically,
\begin{equation}
 J_{2,0}=1,\quad J_{2,2}=3/2,\quad J_{4,0}=3/4,\quad J_{4,-2}=1/2,
\end{equation}
so that the coefficients of the leading O($a^2$) effects can be combined,
\begin{equation}
   t^2{\cal E}_0 = \dfrac{3(N^2-1)}{128\pi^2}\left\{1 + \dfrac{a^2}{t}
    \left(d^{\text{(o)}} - d^{\text{(a)}} - 3d^{\text{(f)}}\right) + \rmO(a^4)\right\}.
\end{equation}
Each $d$-coeffcient takes a particular value e.g. for the Wilson-plaquette, L\"uscher-Weisz or
the clover kernel (or linear combinations thereof). Examples are
\begin{equation}
  d^{\text{(a,o,f)}}  =
  \begin{cases}
               -\frac{1}{24}, & \text{plaquette (pl)},\cr
             \hphantom{+}  \frac{1}{72}, \quad\left( = -\frac{1}{24} -\frac23 c_1\right) & \text{L\"uscher-Weisz (lw)},\cr
             -\frac{5}{24}, & \text{Clover (cl)},
   \end{cases}
\end{equation}
in agreement with~\cite{Fodor:2014cpa}.  Obviously, there are many ways to cancel a single O($a^2$) term, e.g.~by
linear combination of plaquette and clover definitions of the observable, or by a $\tau$-shift~\cite{Cheng:2014jba}.
However, such recipes are of limited use as they only cancel the leading cutoff in a particular observable.
Symanzik improvement is more ambitious in that it aims at improving the action and composite fields by local
counterterms such that leading cutoff effects are eliminated in all observables.
Not so long ago most people (including the authors) would have expected that
tree-level O($a^2$) improvement of flow observables could be achieved by combining an O($a^2$) improved 
action, an O($a^2$) improved observable and the gradient of an O($a^2$) improved action for
the flow (possibly up to a boundary counterterm, cf.~\cite{Sommer:2014mea}).
Hence, using the L\"uscher-Weisz action~\cite{Luscher:1984xn}
for all three kernels should then achieve tree-level O($a^2$) improvement. 
However, the above results show that this choice leads to a non-vanishing coefficient,
\begin{equation}
  d^{\text{total}} = d^{\text{(o)}} - d^{\text{(a)}} - 3d^{\text{(f)}} = -\dfrac{1}{24},
\end{equation}
indicating that this expectation was too naive. It is also worth mentioning that the seemingly
small cutoff effects observed in \cite{Luscher:2010iy} with the clover observable are due
to an accidental cancellation between cutoff effects of clover observable, Wilson flow and Wilson action.

\section{Finite volume: the GF coupling with twisted periodic b.c.'s}

We now pass to a finite volume and consider the gradient flow coupling (\ref{eq:GFcoupling})
with twisted periodic boundary conditions for the gauge field~\cite{Ramos:2014kla}
Since this finite volume scheme preserves translation invariance, the
trace algebra is the same as infinite volume, with the generalized momenta of
the twisted periodic set-up. What changes are the momentum integrals which become momentum sums.
The result can be cast in the same form as eq.~(\ref{eq:oaeffects}) with the
important difference that the numbers $J_{n,m}$ become functions 
of $c=\sqrt{8t}/L$, with the infinite volume result
recovered in the limit $c\rightarrow 0$. As functions of $c$ the different lattice sums are
linearly independent and give thus rise to more stringent improvement conditions, as each of
their coefficients must vanish separately. Our results indicate that tree-level O($a^2$) improvement
cannot be achieved with the L\"uscher-Weisz flow. However, with a generic lattice action for the flow we
found a 1-parameter family of flows such that tree level O($a^2$) improvement is achieved
for this finite volume coupling. Our choice was to set $c_3=0$ and to include
only the bent rectangles/chairs with coefficient $c_2$.
With the choice for the coefficients,
\begin{equation}
   c_0=1, \qquad c_1 =-1/12,\qquad c_2=1/24, 
\end{equation}
this defines what we will call the ``chair flow".

\section{Scaling test in pure SU(3) gauge theory}

We have implemented the chair flow in the open-QCD code~\cite{openQCD}
and performed a scaling test, using the LW action and, for the  observable, the
step-scaling function for the finite volume coupling
with SF boundary conditions and $c=0.3$~\cite{Fritzsch:2013je}. In this case
the expectation value $\langle E(t,x) \rangle $ retains an $x_0$-dependence.
In order to minimize the (small) influence from the time boundaries
we set $x_0=T/2$ and $T=L$. Moreover, the (colour-) magnetic and
electric components of $G_{\mu\nu}$ contribute differently. After some experimentation
with perturbative data we chose to only use the magnetic components, i.e.
\begin{equation}
-\frac12 \langle \tr{G_{kl}(t,x)G_{kl}(t,x)}\rangle \vert_{x_0=T/2} = {\cal N}(c,a/L) \bar{g}_{\rm GF}^2(L).
\end{equation}
We then computed the step-scaling function $\Sigma(u,a/L)=\bar{g}_{\rm GF}^2(2L)$
at $u=\bar{g}_{\rm GF}^2(L)=2.6057$ for lattice sizes $L/a = 8,12,16,24$ and with scale factor $s=2$.
The results shown in figure~1 show indeed a strong reduction of cutoff effects for
the improved data with the chair flow. On the coarsest lattice the cutoff effects
are reduced by a factor 4 or so with respect to the second best definition with
the Wilson flow and the clover observable.

\begin{figure}
\includegraphics[width=0.45\textwidth]{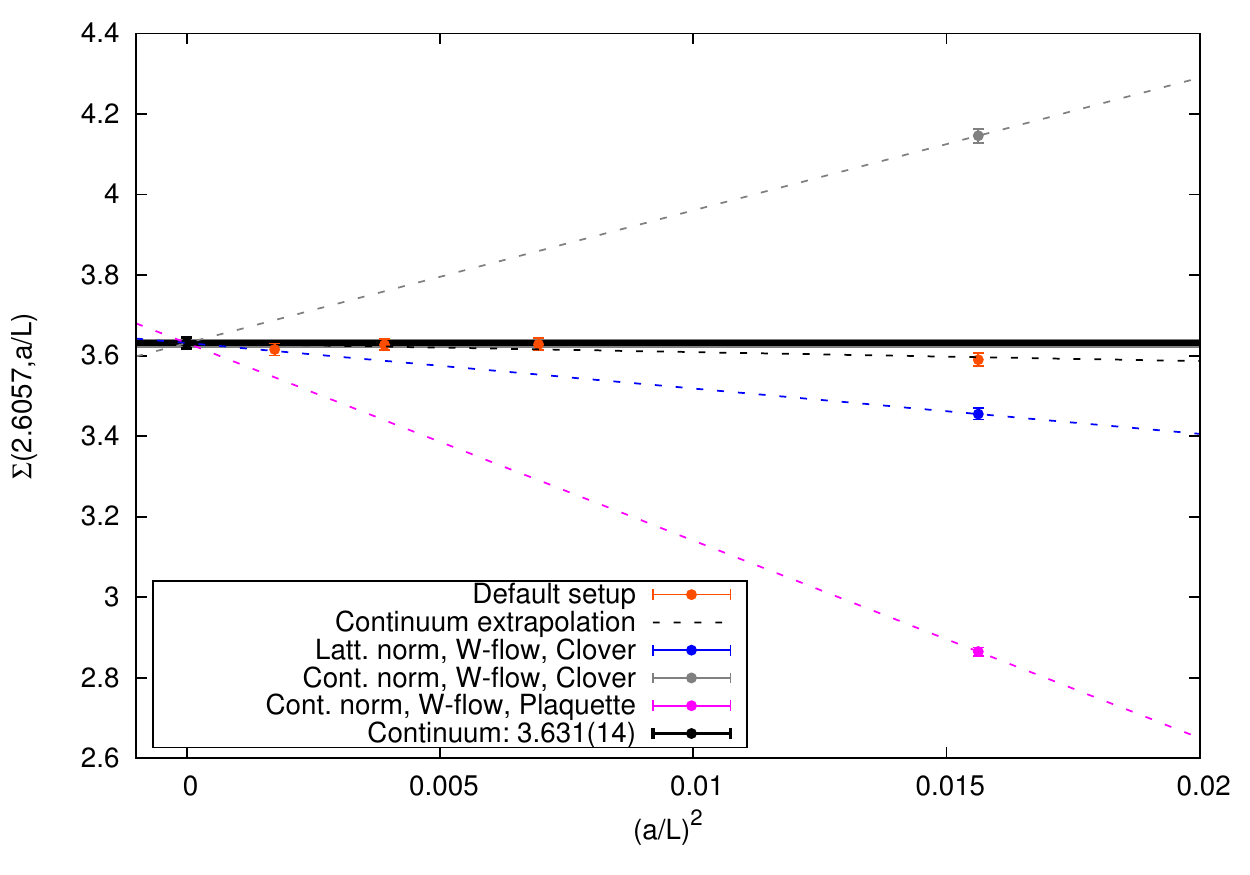}
\includegraphics[width=0.45\textwidth]{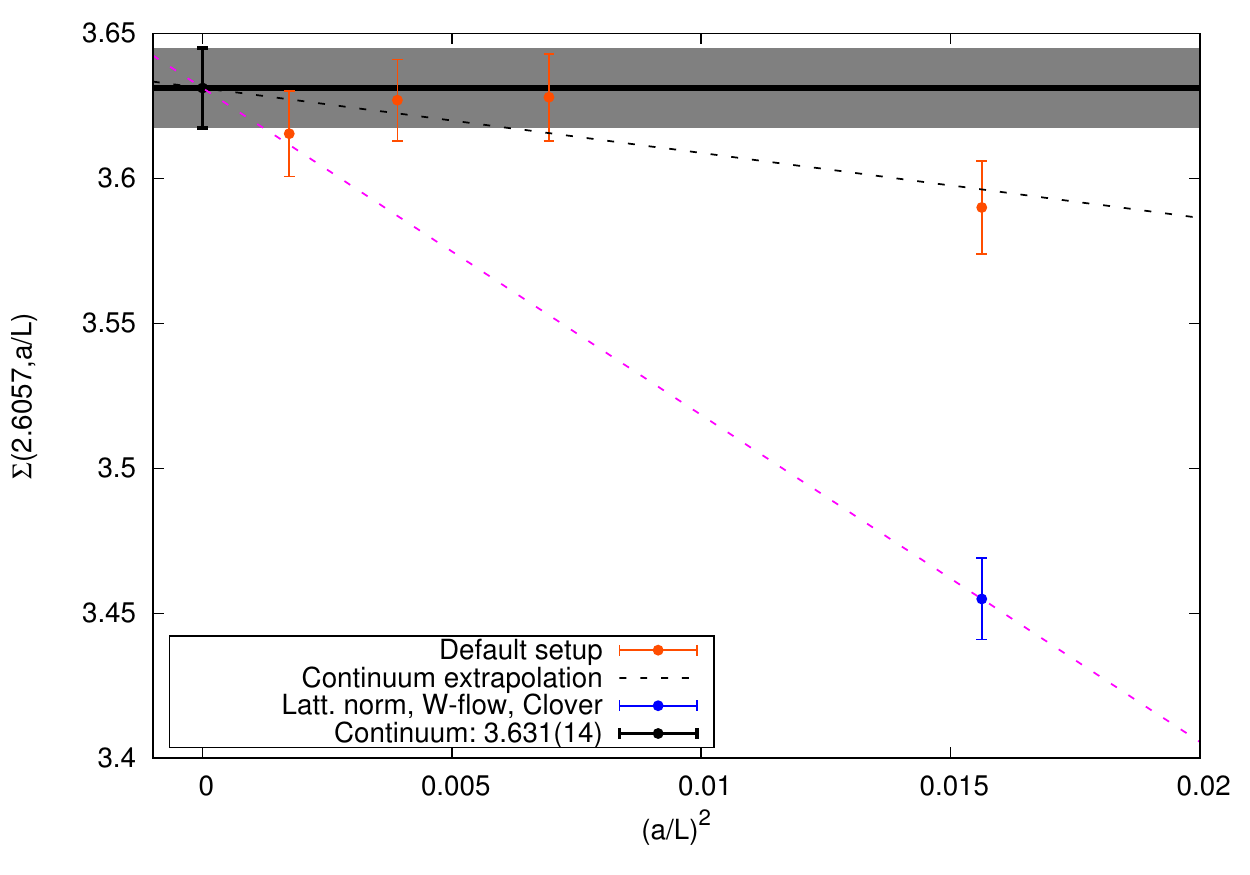}
\caption{The step-scaling function computed with the ``default" set-up using an improved observable, action and the chair flow.
For comparison we have included several data points at the coarsest lattice spacing with alternative discretizations.
The definition with the clover observable and the Wilson flow fares second best and is shown again in the right panel.}
\label{fig1}
\end{figure}

\section{Symanzik O($a^2$) improvement of flow observables}

Despite the impressive reduction of lattice artefacts seen with the chair flow it
remains unclear whether this set-up is improved in the
sense of Symanzik~\cite{Symanzik:1983dc,Symanzik:1983gh,Luscher:1984xn}.
To apply the Symanzik programme to gradient
flow observables it is essential to use a local formulation of the theory.
This is achieved by the $4+1$ dimensional set-up discussed
by L\"uscher and Weisz~\cite{Luscher:2011bx}, with the flow time $t$ as additional coordinate.
The $4+1$ dimensional lattice action is 
\begin{equation}
S = S_{\rm lat}[U] - 2 \int_0^\infty\rmd t\, a^4 \sum_x \tr\biggl(
L_\mu(t,x) \left\{a^{-1}\left(\partial_t V_\mu(t,x)\right)V_\mu(t,x)^\dagger
 + a^{-3}\partial_{x,\mu}\left(g_0^2 S_{\rm lat}[V]\right)\right\}\biggr).
\end{equation}
The continuum limit of this action is the starting point for the Symanzik expansion.
O($a^2$) counterterms could arise from 3 sources, namely the observable, the 4-dimensional boundary
at $t=0$, and O($a^2$) counterterms in the $4+1$-dimensional bulk.
Following the reasoning by L\"uscher in the fermionic case~\cite{Luscher:2013cpa}, we expect
classical O($a^2$) improvement for both the observable and the flow to {\em completely} remove
O($a^2$) effects from these sources, i.e.~to all orders in the coupling!
This leaves us with possible boundary effects,
generated by dimension 6 terms at $t=0$. If these do not involve the field $L_\mu(t,x)$ they
must be of the same form as the standard improvement counterterms in the 4-dimensional
theory. Requiring improvement for non-flow observables fixes this freedom.
After use of the flow equation we are thus left with the following 
candidate counterterms\footnote{We thank A.~Patella for pointing
out to us the possibility of a term quadratic in $L_\mu$.},
\begin{equation}
   \tr\{ L_\mu(t,x) L_\mu(t,x)\}\vert_{t=0},\qquad \tr\{ L_\mu(t,x) D_\nu G_{\nu\mu}(t,x)\}\vert_{t=0},
\end{equation}
which may or may not be required for O($a^2$) improvement.
How to best implement these counterterms and whether both are really required is left to future work. 
Our explicit calculations suggest that they do not contribute at tree-level of perturbation theory.

\section{Classical expansion of the flow equation}

It remains to carry out the classical $a^2$-expansion of the lattice flow equation, where
$V_\mu(t,x)$ is related by a path ordered exponential to the continuum gauge field $B_\mu$.
We performed this calculation with the gradient of a lattice action with free parameters $c_0,c_1$ and $c_2$.
Unfortunately it turns out that the chair flow does not seem to be O($a^2$) improved.
During the calculation we however noticed that the O($a^2$) effects for the
L\"uscher-Weisz gradient flow ($c_0=5/3$, $c_1=-1/12$, $c_2=c_3=0$)
have a simple structure:
\begin{equation}
  \partial_t B_\mu = D_\nu G_{\nu\mu}-\frac{1}{12} a^2 D_\mu^2D_\nu G_{\nu\mu} + O(a^3).
\end{equation}
This suggests a simple  modification of the lattice flow equation,
\begin{equation}
  a^2\left(\partial_t V_\mu(t,x)\right)V_\mu(t,x)^\dagger =
  -\left(1+\frac{1}{12}a^2\nabla_\mu^\ast\nabla_\mu^{}\right)\partial_{x,\mu}\left(g_0^2 S_{\rm lat}[V]\right),
\end{equation}
where $\nabla_\mu$ and $\nabla_\mu^\ast$ denote the covariant adjoint lattice derivative operators.
While this "Zeuthen flow" removes all O($a^2$) effects from the flow equation, it still remains to be
put to a numerical test.

\section{Conclusions}

We have carried out a detailed investigation of tree-level O($a^2$) effects in gradient flow observables derived
from $E(t,x)$.
This led to the definition of the chair flow, which eliminates O($a^2$) tree-level effects from the flow equation,
in all cases considered. Furthermore, a quenched scaling test showed remaining cutoff effects to be very small.
In order to see whether this provides a complete solution for {\em any} flow observable we took
to Symanzik O($a^2$) improvement in the $4+1$ dimensional set-up: a couple of candidate counterterm remain
to be investigated, but do not seem to contribute in the explicit tree-level calculations.
Symanzik improvement requires the classical $a^2$-expansion of both observables and flow equation.
It turns out that the chair flow is not fully O($a^2$) improved, i.e.~the cancellation seen in the explicit
calculation might be specific to the observables considered.
Finally we proposed the fully O($a^2$) improved "Zeuthen flow" as a simple modification of the L\"uscher-Weisz gradient flow,
which still needs to be tested numerically.

\begin{center}
{\bf Acknowledgments}
\end{center}
We are grateful to our colleagues in the ALPHA collaboration for discussions and
an enjoyable collaboration. S.S. acknowledges support by SFI under grant 11/RFP/PHY3218.
We thank ICHEC and DESY-Zeuthen for providing computer resources for the numerical simulations.

\end{document}